\documentclass[prd,superscriptaddress,amsmath,amssymb,onecolumn,floatfix]{revtex4}
\usepackage{setspace}
\usepackage{graphicx}
\usepackage{hyperref}

\begin{document}
\begin{flushleft}
KCL-PH-TH/2022-04
\end{flushleft}

\title{Gravitational Waves: the theorist's  Swiss knife}

\author{Mairi Sakellariadou }
\affiliation{Theoretical Particle Physics and Cosmology Group, \, Physics \, Department, \\ King's College London, \, University \, of London, \, Strand, \, London \, WC2R \, 2LS, \, UK}

\date{\today}

\begin{abstract}
Gravitational waves provide a novel and powerful way to test astrophysical models of compact objects, early universe processes, beyond the Standard Model particle physics, dark matter candidates, Einstein's theory of General Relativity and extended gravity models, and even quantum gravity candidate theories. A short introduction to the gravitational-wave background and the method we are using to detect it will be presented. Constraints on various astrophysical/cosmological models from  the non-detectability of the gravitational-wave background will be discussed. Gravitational waves from transients will be highlighted and their physical implications will be summarised.

\end{abstract}

\maketitle


\section{Introduction}

The first direct detection of gravitational waves (GWs) from the LIGO interferometers \cite{LIGOScientific:2016aoc}, followed by tens of other events \cite{LIGOScientific:2018mvr,LIGOScientific:2020ibl,LIGOScientific:2021usb}, opened a new window into the early and late universe and offered a powerful tool to test theories that are otherwise beyond the current experimental realm \cite{Barausse:2020rsu}. Gravitational waves provide valuable information on astrophysical models of compact objects \cite{LIGOScientific:2020kqk,LIGOScientific:2021psn,Martinovic:2021fzj}, the cosmic history of the universe \cite{LIGOScientific:2019zcs,LIGOScientific:2021aug}, and the large-scale structure  \cite{Bellomo:2021mer} independently of electromagnetic waves and corresponding observational tools like the cosmic microwave background. In addition, gravitational waves offer the means to test early universe processes \cite{Martinovic:2021hzy,Romero-Rodriguez:2021aws,Badger:2021enh}, beyond the Standard Model  particle physics at energy scales that cannot be reached by current or near-future particle accelerators \cite{LIGOScientific:2021nrg,Romero:2021kby}, dark matter candidates \cite{Sagunski:2017nzb,Huang:2018pbu,Zhang:2021mks,LIGOScientific:2021jlr,LIGOScientific:2021odm}, Einstein's theory of General Relativity \cite{LIGOScientific:2020tif,LIGOScientific:2021sio}, modified/extended gravity proposals \cite{Sagunski:2017nzb,LISACosmologyWorkingGroup:2019mwx,Lambiase:2020vul,Martinovic:2019hpo}, and even quantum gravity candidate theories \cite{Calcagni:2019kzo,Calcagni:2019ngc,Addazi:2021xuf,Jenkins:2018ysa}. 

The LIGO/Virgo interferometers have so far detected several tens of phase-coherent transient signals emitted by compact binaries (black holes, neutron stars, or combination of both). In addition, there is an ongoing search for phase-coherent persistent signals, like the continuous waves emitting by spinning neutron stars 
\cite{LIGOScientific:2022pjk,LIGOScientific:2021ozr,LIGOScientific:2020qhb}, and phase-incoherent transient or persistent signals; the latter one being the long-sought stochastic gravitational-wave background. The techniques used to search for  GW signals depend on their nature  and it is important to note that even in the case of a non-detection (like for the gravitational-wave background), still important constraints can be imposed to various theoretical models and their free parameters.

In what follows, I will briefly summarise the current status of constraints imposed to various astrophysical and cosmological models, as well to fundamental theories using the novel and powerful tool of gravitational waves.

\section{Gravitational-wave background}

Our universe is permeated by cosmic microwave background electromagnetic radiation, the relic photons of the Big Bang, first discovered by Penzias and Wilson in 1965.  Similarly, our universe is expected to be filled with a stochastic gravitational-wave background generated by various processes that took place in its early history. In addition, a background of gravitational waves can also emerge from the incoherent superposition of a large number of astrophysical sources, which are too weak to be detected separately, and such that the number of sources that contribute to each frequency bin is much larger than one. These two components (the stochastic cosmological and the astrophysical one) form the gravitational-wave background (GWB).

The GWB \cite{Romano:2016dpx} can be expressed in terms of a dimensionless parameter (like all cosmological density parameters), which expresses the energy density of gravitational waves $\rho_{\rm gw}$ in a logarithmic interval of frequency $f$, normalised by the critical energy density $\rho_{\rm cr}=3H_0^2c^2/(8\pi G)$, as
\begin{equation}
  \Omega_{\rm gw}(f)=\frac{f}{ \rho_{\rm cr}}\frac{{\rm d}\rho_{\rm gw}}{ {\rm d} f}~,
\end{equation}
with $\rho_{\rm gw}\sim \dot h^2$, where $h$ denotes the strain of GWs and overdot stands for derivative with respect to cosmological time $t$.

The GWB is characterised by random non-deterministic phase evolution of GWs emitted from a large number of distant sources. On the one hand, the GWB of cosmological origin looses any phase coherence present at emission or at horizon crossing, as metric perturbations affect the phase of GWs propagating through the inhomogeneous universe.
On the other hand, the GWB of astrophysical origin, generated by overlapping signals from many individual sources below confusion limit of detector, has no phase-coherence. 
For a standard cosmological model, it has been shown \cite{Margalit:2020sxp} that there is complete phase decoherence of the GWB above frequencies $f\sim 10^{-12}$ Hz. Hence, only the amplitude of GWs plays a r\^ole in estimating $\Omega_{\rm gw}$:
\begin{equation}
    \Omega_{\rm gw}\sim\langle{\tilde h}{\tilde h}^\star \rangle = {\cal A}^2,
\end{equation}
where ${\tilde h}$ denotes the strain in Fourier space and ${\cal A}$ stands for the GW amplitude.

A detection of the GWB from unresolved compact binary coalescences (CBC) could be made by Advanced LIGO and Advanced Virgo at their design sensitivities.
It would appear as noise in a single GW detector $i$:
\begin{equation}
    {\tilde s}_i(f)={\tilde h}_i(f) +{\tilde n}_i(f)~,
\end{equation}
with the noise $n$ being much bigger than the strain $h$ at a given frequency $f$. Thus, to detect the GWB, we need to take the correlation between two detector outputs $i, j$:
\begin{equation}
    \langle{{\tilde s}^\star_i(f) {\tilde s}_j}(f')
    \rangle
    =
    \langle{{\tilde h}^\star_i(f) {\tilde h}_j}(f')
    \rangle
    +
    \langle{{\tilde h}^\star_i(f) {\tilde n}_j}(f')
    \rangle
    +\langle{{\tilde n}^\star_i(f) {\tilde h}_j}(f')
    \rangle
    +\langle{{\tilde n}^\star_i(f) {\tilde n}_j}(f')
    \rangle~.
\end{equation}
The signal-to-noise (SNR) ratio grows slowly over observing time $T_{\rm obs}$:
\begin{equation}
    \langle s_1 s_2\rangle \sim {\rm Var}[s_1s_2]\sim T_{\rm obs}\Rightarrow {\rm SNR}=\frac{\langle s_1 s_2\rangle}{\sqrt{{\rm Var}[s_ss_2]}}\sim \sqrt{T_{\rm obs}}~.
\end{equation}
Assume that the GWB is isotropic, Gaussian, stationary and unpolarised:
\begin{equation}
    \langle {\tilde h}_A(f,{\hat n)}
    {\tilde h}_{A'}^\star(f',{\hat n')}\rangle
    =\frac{3H_0^2}{32\pi^3 f^3}
    \Omega_{\rm gw}(f)\delta_{AA'}\delta (f-f')\delta^2({\hat n}, {\hat n}')~,
\end{equation}
where $H_0$ stands for the Hubble constant and $A, A'$ denote the $+, x$ polarisation modes. We  build the estimator ${\hat C}_{ij} (f;t)$ for the GWB measured from detectors $i, j$:
\begin{equation}
  {\hat C}_{ij} (f;t)=
  \frac{2}{T}\frac{{\rm Re}[{\tilde s}^\star_i(f;t){\tilde s}^\star_j(f;t)]}{\Gamma_{i j}(f) S_0(f)}~,
\end{equation}
where ${\tilde s}_i(f;t)$ denotes the Fourier transform of the strain time series in detector $i$ starting at time $t$,  $\Gamma_{i j}(f)$ is the normalised overlap reduction function between detectors $i$ and$j$ accounting for all the geometric factors that come into play when cross-correlating data from different detectors, $T$ stands for the duration over which the Fourier transform is taken, and $S_0(f)=3H_0^2/(10\pi^2f^3)$ is the spectral shape for a GWB that is flat in energy density. Since the GWB is considered to be isotropic, Gaussian, stationary and unpolarised,  it is well-described by a single power spectral density $S_{\rm gw}(f)$:
\begin{equation}
   \langle {\tilde h}_i^\star(f){\tilde h}_j(f')\rangle 
   =
   \frac{1}{2}\delta _T(f-f')\Gamma _{ij}(f)S_{\rm gw}(f)~,
\end{equation}
where $\delta_T(f-f')$ denotes the finite-time approximation to the dirac delta function and $S_{\rm gw}(f)$ is related to $\Omega_{\rm gw}(f)$ through
\begin{equation}
    S_{\rm gw}(f)=\frac{3H_0^2}{10\pi^2}\frac{\Omega_{\rm gw}(f)}{f^3}~.
\end{equation}
Assuming the GW signal and the intrinsic noise to be uncorrelated
and the noise in each frequency bin to be independent, we obtain
\begin{equation}
\langle{\hat C}_{ij} (f;t)\rangle=
\Omega_{\rm gw}(f)+
2{\rm Re}\left[
\frac{\langle {\tilde n}_i^\star(f;t){\tilde n}_j(f;t)\rangle}{T\Gamma_{ij}(f)S_0(f)}
\right]~.
\end{equation}
In the absence of correlated noise, $\langle {\tilde n}_i^\star(f){\tilde n}_j(f)\rangle=0$, the 
$\langle{\hat C}_{ij} (f)\rangle$ is an estimator of the dimensioneless gravitational wave energy density $\Omega_{\rm gw}(f)$. However, it may be that $\langle {\tilde n}_i^\star(f){\tilde n}_j(f)\rangle\neq 0$;
a potential source of correlated magnetic noise is Schumann resonances. 

Schumann resonances are global extremely low frequency peaks in the electromagnetic field of the Earth. They are eigenmodes of the conducting spherical cavity formed by the surface of the Earth and its ionosphere, and are excited by lightning discharges. 
These low-frequency resonances in the Earth’s global electromagnetic field could couple to the mass suspension system and electronics in the interferometers, and mimic a GWB signal.
One must therefore develop a method for detecting correlated magnetic noise and separating from a true GWB signal. This issue is usually addressed by performing a coherent subtraction in the raw data using Wiener filtering. However,  for weakly coupled signals it can be difficult to completely subtract the noise. Recently, an alternative novel method to minimise the possibility of a false detection was proposed and tested with realistic simulations of correlated magnetic noise in interferometric GW detectors in \cite{Meyers:2020qrb}. 

The correlated magnetic noise in GW detectors was modelled using data collected by magnetometers placed near the detector sites, and adopting a parameterised model for the magnetic field to GW detector coupling \cite{Meyers:2020qrb}. 
Peaks in the detectors’ strain channels were related to the peaks in the magnetometer channels via a coupling function, assumed to be constant in time,  well-described by a power-law, with index $\beta$.
Hence,
\begin{equation}
    \langle{\hat C}_{ij} (f)\rangle=\Omega_{\rm gw}(f)+\Omega_{{\rm M}, ij}(f)~,
\end{equation}
with $\Omega_{{\rm M}, ij}(f)$ representing the magnetic contribution:
\begin{equation}
    \Omega_{{\rm M}, ij}(f)=
    \kappa_i\kappa_j\left(\frac{f}{10 {\rm Hz}}
    \right)^{-\beta_i-\beta_j}
    {\hat M}_{ij}(f)\times 10^{-22}~;
\end{equation}
$ \kappa$ denotes the amplitude of the coupling at 10 Hz at the site of interferometric GW detector.
The factor $10^{-22}$ assumes that the unites of magnetometer data ${\hat m}_i(f)$ are T/Hz.

Bayesian model selection was then used to distinguish between models that include correlated magnetic noise and those with a true GWB. This method originally proposed \cite{Meyers:2020qrb} and used in the framework of LIGO/Virgo detectors \cite{KAGRA:2021kbb}, was consequently extended in the case of the 3g  Einstein Telescope interferometer \cite{Janssens:2021cta} forecasting the necessary measures to ensure that magnetic contamination will not pose a threat to the corresponding science goals.  In particular, it was shown \cite{Janssens:2021cta} that for GWB searches below $\sim 30$ Hz it will be necessary for the Einstein Telescope magnetic isolation coupling to be two to four orders of magnitude better than the one measured in the current Advanced LIGO and Virgo interferometers.

In the Advanced LIGO's and Adavanced Virgo's third observing run (O3) data from all three baselines were included in the isotropic GWB search. Combining O3 data with upper limits from the earlier first and second  (O1 and O2) observing runs, it was found that the results are consistent with uncorrelated noise, hence placing upper limits on the strength of the GWB \cite{KAGRA:2021kbb}. More precisely, performing a Bayesian analysis that allows for the presence of both a GWB and an effective magnetic background arising from Schumann resonances, no evidence of correlated noise of magnetic origin was found.

Considering a power-law GWB of the form
\begin{equation}
    \Omega_{\rm gw}(f)=\Omega_{\rm ref}\left(\frac{f}{f_{\rm ref}} \right)^\alpha~,
\end{equation}
where $f_{\rm ref}$ is a fixed reference frequency chosen to be 
$f_{\rm ref}=25$ Hz and $\alpha$ is the spectral index describing the source of the GWB, limits on the dimensionless energy density in GWs are imposed for 
$\alpha=0, 2/3, 3$.

The $\alpha=0$
approximately characterises a GWB originated by a network of cosmic strings, or a period of slow-roll inflation in the LIGO-Virgo frequency band.
The $\alpha= 2/3$ power-law describes the GWB from compact binaries coalescences when contributions from the inspiral phase dominate the GWB.
This can be easily seen by writing down the quadrupole expression for the energy $E_{\rm gw}$ released per frequency from the inspiral phase of two compact objects of masses $m_1, m_2$:
\begin{equation}
\frac{{\rm d}E_{\rm gw}}{{\rm d}f}=\frac{(G\pi)^{2/3}}{3}\frac{m_1m_2}{(m_1+m_2)^{1/3}}f^{-1/3}~,
\end{equation}
and then integrating over all frequencies $f$.
While this consideration is a very good approximation in the LIGO-Virgo frequency band, it is not be valid for mergers of binaries arising from population III stars \cite{Martinovic:2021fzj}. The $\alpha=3$ corresponds to a GWB that is flat in the strain power, $S_h(f)\propto f^{-3}\Omega_{\rm gw}(f)$ and describes a GWB from supernovae. 

The GWB searches from the O1, O2 and O3 observing runs led to the following upper limits on the strength of the GWB \cite{KAGRA:2021kbb}:
\begin{itemize}
\item $\Omega_{\rm gw}\leq 5.8\times 10^{-9}$ at the $95\%$ credible level for frequency-independent $(\alpha =0)$ GWB, using a uniform prior in the log of the strength of the GWB, with $99\%$ of the sensitivity coming from the band 20-76.6 Hz;
\item
$\Omega_{\rm gw}\leq 3.4\times 10^{-9}$ at $f_{\rm ref} =25$ Hz for $\alpha= 2/3$ power-law GWB in the band 20-90.6 Hz;
\item
$\Omega_{\rm gw}\leq 3.9\times 10^{-10}$ at $f_{\rm ref} =25$ Hz for $\alpha= 3$ power-law GWB in the band 20-291.6 Hz.
\end{itemize}
Detection of a GWB is indeed hard, as one can easily observe by comparing the dimensionless GW energy parameter to the corresponding of the cosmic microwave background ($\Omega_{\rm cmb}$): $\Omega_{\rm gw}\ll \Omega_{\rm cmb}\approx 10^{-5}$.

 The challenge of claiming a true detection of the GWB will be immediately succeeded by the difficulty of relating the signal to the sources (astrophysical or cosmological) that contribute to it. This task was addressed in \cite{Martinovic:2020hru}, considering that compact binary coalescences and two additional cosmological sources, namely cosmic strings (CS) and early universe first order phase transition (FOPT), comprise the GWB. It was shown \cite{Martinovic:2020hru} that the network of Advanced LIGO and Advanced Virgo, operating at design sensitivity, will not be able to allow for separation of these sources. However, 3g detectors (such as Einstein Telescope and Cosmic Explorer),  which are sensitive to most individual compact binary mergers, may reduce the astrophysical signal via subtraction of individual sources, and potentially reveal a cosmological background.
 Performing a Bayesian analysis, it was found \cite{Martinovic:2020hru} that considering an Einstein Telescope and Cosmic Explorer network and reasonable levels of individual source subtraction, will allow us to detect a cosmological signal of $\Omega_{\rm cs}(25 {\rm Hz})\simeq4.5\times 10^{-13}$    for cosmic strings and $\Omega_{\rm bpl}(25 {\rm Hz})\simeq 2.2\times 10^{-13}$    for a broken power-law model describing GWs produced during an early universe phase transition at $f_{\rm ref}= 25$ Hz.

\subsection{GWB from CBC: information on compact binaries}
 
Consider an astrophysical GWB composed by compact binaries of population I and II (pop I/II) stars. The energy density of GWs from compact binaries as a function of frequency $f$ and model parameters $\theta$ can be expressed as
\begin{equation}
\Omega_{\rm gw}(f; \theta) =\frac{f}{\rho_{\rm cr} H_0}
\int {\rm d}\theta p(\theta)
\int_0^{z_{\rm max}(\theta)} \frac {R_{\rm m}(z;\theta)
\frac{{\rm d}E_{\rm gw}(f_{\rm s};\theta)}{{\rm d}f_{\rm s}}
}
{(1+z)E_z(\Omega_{\rm m},\Omega_\Lambda, z)}{\rm d}z,
\end{equation}
where $p(\theta)$ is the probability distribution of the source parameters, ${\rm d}E_{\rm gw}(f_{\rm s};\theta)/{\rm d}f_{\rm s}$ is the energy density emitted by as single source at a redshift $z$ with parameters $\theta$, $E(\Omega_{\rm m},\Omega_\Lambda, z)=\sqrt{\Omega_{\rm m}(1+z)^3+\Omega_\Lambda}$ accounts for the expansion history of the universe  with 
$\Omega_{\rm m}=0.31$ and $\Omega_\Lambda=0.69$ the matter and dark energy contribution to the dimensioneless cosmological energy density, respectively, $f_{\rm s}=(1+z)f$ is the emitted frequency in the source frame, and $z_{\rm max}(\theta)$ is the maximal redshift at which a compact binary with parameters $\theta$ can form. 

The most important quantities describing binary black holes ate the masses and spins of each black hole component. The impact on uncertainties in the distribution of binary black holes on the astrophysical GWB has been investigated in \cite{Jenkins:2018kxc} using population inference from GW observations. Adopting a truncated power-law black hole mass distribution
and modelling the black hole spin magnitudes by a Beta distribution, 
about 10,000 models have been considered, showing that the astrophysical GWB monopole is sensitive to the nature of the binary black hole population, and in particular the local merger rate. Hence, the GWB can be used as a new probe of the population of compact objects throughout cosmic history, complementary to direct observations.

The theoretically postulated population III (pop III) stars are thought to have formed at high redshifts, hence they have low metallicity as compared to pop I/II stars which are more recently formed.
If pop III exists, their GWB signal will be superposed with a pop I/II signal. Several models indicate that the total CBC background is dominated by pop I/II, implying that the pop III signal will remain hidden underneath the pop I/II signal. For second generation detector networks (even by including LIGO-Hanford, LIGO-Livingston, Virgo, LIGO- India, and KAGRA) the pop III signal is lost in the pop I/II foreground and hence invisible.
However, third generation  terrestrial detectors, such as Einstein Telescope and Cosmic Explorer, may reveal a pop III background \cite{Martinovic:2021fzj}.

A CBC signal is typically modelled with a 2/3 spectrum. This is appropriate for pop I/II stars which lead to a GWB in the LIGO/Virgo frequency star at the inspiral phase. However, the older pop III stars would show up at the later merger and ringdown phases, hence there will be a clear deviation from the 2/3 spectrum. Testing filters different from a 2/3 power spectrum and performing a Bayesian analysis, it was shown \cite{Martinovic:2021fzj} that a broken power-law filter is the most appropriate one to reveal a pop III GWB. 

Considering a detector network composed by Einstein Telescope and two Cosmic Explorers, it was shown that one could successfully subtract enough individual merger events to unravel the elusive pop III GWB. The reason being that subtraction methods are more effective for nearby sources, so the residual pop I/II signal can become sub-dominant to the pop III residual. 
It was also found that
a detection of a pop III GWB could reveal important information, such as the mass and redshift distribution of pop III stars \cite{Martinovic:2021fzj}.

\subsection{Stochastic GWB from CS: information on particle physics models beyond the Standard Model}

Cosmic strings are one-dimensional topological defects, which may have been formed in the early universe as a result of a phase transitions followed by spontaneously broken symmetries, characterised by a vacuum manifold with non-contractible closed curves. These defects have been shown to be generically formed in the context of Grand Unified Theories \cite{Jeannerot:2003qv}.

A cosmic string network reaches an attractor scaling solution in which its energy density remains a fixed fraction of the background energy density \cite{Bennett:1987vf,Sakellariadou:1990nd,Ringeval:2005kr}. As  strings collide, they always intercommute exchanging partners and reconnect after a collision. As a result, closed loops are formed when a string self-intersects or two curved strings collide. Loops smaller than the horizon decouple from the cosmological evolution and oscillate under their own tension, slowly decaying into GWs \cite{Vachaspati:1984gt}.

The relativistic nature of strings leads to the formation of points where the string momentarily moves at the speed of light; these points are called cusps. The intersections of strings generate discontinuities on their tangent vecto; they are called kinks. Cusps and kinks generate gravitational wave bursts \cite{Damour:2000wa}.
Cusps produce beamed GWs in the forward direction of the cusp, kinks propagate around the string creating GWs with a fan-like emission, while the collision of two kinks radiates GWs isotropically. The high-frequency gravitational-wave spectrum of an oscillating loop is dominated by bursts emitted by cusps and kinks. 
The waveforms of the strain are power-law functions in the frequency domain, and grow as $f^{-4/3}, f^{-5/3}$ $f^{-2}$ for  cusp, kink, and kink-kink collision cases, respectively.
In addition to these distinct individual bursts, the incoherent superposition of weaker GW bursts from cosmic strings produced over the history of the universe leads to a strongly non-Gaussian stochastic GWB. 

The cosmic string network also contains long strings that stretch across a Hubble volume, which are either infinite or in the form of super-horizon loops; they also emit GWs \cite{Sakellariadou:1990ne}. However, the dominant  GW contribution is the one produced by the superposition of GW radiation from  sub-horizon loops along each line of sight.

The spectrum of the GWB from cosmic strings is
\begin{equation}
    \Omega_{\rm gw}(f)=\frac{4\pi^2}{3H_0^2}f^3\sum_i\int {\rm d}z\int {\rm d}\ell h_i^2\times \frac{{\rm d}^2 R_i}{{\rm d}z{\rm d}\ell}~,
\end{equation}
where ${\rm d}^2 R_i/{\rm d}z/{\rm d}\ell$ is the burst rate of type $i$ -- cusp, kink or kink-kink collision -- per unit loop size $\ell$ and per redshift $z$; it depends on the number of cusps and kinks.

Consider Nambu-Goto strings, parametrised by a single quantity, namely the string tension $G\mu$ (with $G$ Newton's constant and $\mu$ the string linear mass density; note that we set $c=1$), related to the energy scale ${\bar \eta}$ at string formation by
$G\mu\sim ({\bar\eta}/M_{\rm Pl})^2$ (with $M_{\rm Pl}$ the Planck mass). 
To estimate the GWB generated by cosmic strings one must know the loop distribution and the average number of cusps and kinks.
Numerical simulations have shown that while kinks
accumulate over the cosmological evolution, the number of cusps per loop remains yet undetermined.

There are two different models \cite{Lorenz:2010sm,Blanco-Pillado:2013qja} for the loop distribution based on different numerical simulations. Their differences arise from different modelling of the production and cascade of loops from the infinite string network. In addition, a new agnostic model that extends and encompasses the ones based on the numerical simulations, has been recently proposed and studied \cite{Auclair:2019zoz}.

A search for a stochastic GWB was conducted using the LIGO and Virgo O3 data and combining them with those from the previous O1 and O2 runs.
Performing a Bayesian analysis taking into account the precise shape of the GWB from cosmic strings, upper limits were derived on the cosmic string tension \cite{LIGOScientific:2021nrg}.
More precisely, for model \cite{Blanco-Pillado:2013qja} the stochastic analysis ruled out particle physics models leading to cosmic strings with tension $G\mu\gtrsim (9.6\times 10^{-9}-10^{-6})$. 
For this model, the strongest limit comes from pulsar timing measurements, excluding string tension $G\mu\gtrsim 10^{-10}$.
For model \cite{Lorenz:2010sm}, the stochastic analysis ruled out particle physics models with string tension
$G\mu\gtrsim (4.0-6.3)\times10^{-15}$; this is the strongest limit  in the literature. Two different variants of model \cite{Auclair:2019zoz} were considered. One that reproduces qualitatively the loop production function of model \cite{Blanco-Pillado:2013qja}   in the radiation era and the loop production of model \cite{Lorenz:2010sm} in the matter era, for which string tension $G\mu\gtrsim(2.1-4.5)\times 10^{-15}$ was ruled out aside form a small region where the number of kinks is greater than 180. 
For this variant, the exclusion range obtained from the stochastic LIGO/Virgo analysis is the strongest one in the literature.
The other variant reproduces qualitatively the loop production function of model \cite{Lorenz:2010sm} in the radiation era and the loop production of model \cite{Blanco-Pillado:2013qja} in the matter era, for which particle physics models with string tension $G\mu\gtrsim (4.2-7.0)\times 10^{-15}$ were excluded.
The corresponding constraint obtained from the cosmic microwave background is almost as strong as this one.
We note that the average number of cusps per oscillation on a cosmic string loop has been set to 1. A high number of cusps leads qualitatively to a similar result as increasing the number of kinks: for model \cite{Blanco-Pillado:2013qja}, the 
constraints are weakened, whereas for all other considered models the bounds are insensitive to the number of cusps \cite{LIGOScientific:2021nrg}.

 Considering the cosmic string models  \cite{Lorenz:2010sm,Blanco-Pillado:2013qja} for the loop distribution, it was shown \cite{Auclair:2019wcv} that LISA will be 
 able to probe cosmic strings with tension $G\mu\gtrsim {\cal O}(10^{-17})$, improving by about 6 orders of magnitude current pulsar timing arrays (PTA) constraints, and potentially 3 orders of magnitude with respect to expected constraints from next generation PTA observatories. However, to obtain these limits only the LISA noise was taken into account.  
 Considering the cosmic string GWB in the presence of a compact binary produced astrophysical background and a galactic foreground, in addition to the LISA noise,  it was shown  \cite{Boileau:2021gbr} that with  four years of data, LISA will be able to measure a cosmic string tension $G\mu\approx 10^{-16}$ (for the model \cite{Kibble:1984hp}) to $G\mu\approx 10^{-15}$  (for the model \cite{Ringeval:2005kr,Lorenz:2010sm,Blanco-Pillado:2013qja}) or bigger; the galactic foreground affecting the $G\mu$ limit more than the astrophysical background.

\subsection{Stochastic GWB from FOPT: information on particle physics beyond the Standard Model}

As the universe expands and its temperature drops, a series of phase transitions might have taken place. In the case of a first-order phase transition (FOPT), once the temperature drops below a critical value, the universe, being originally in a meta-stable phase, reaches a stable one, through a sequence of bubble nucleation, growth, and merger. During this process, a stochastic GWB is thought to be generated \cite{Witten:1984rs,Hogan:1986qda} by mainly three sources: bubble collisions, sound waves, and magnetohydrodynamic turbulence. Several extensions of the Standard Model predict strong FOPTs; their footprint on the stochastic GWB may thus reveal important information on beyond the Standard Model  particle physics constraining their free parameters.

The stochastic GWB produced by a FOPT covers a large frequency range, with the peak frequency set by the temperature $T_{\rm FOPT}$ at which the FOPT takes place. For a $T_{\rm FOPT}\sim (10^7-10^{10})$ GeV, which is the case for the Peccei-Quinn axion model or high-scale supersymmetry models, the generated  stochastic GWB lies within the frequency range of Advanced LIGO and Advanced Virgo. Thus, current terrestrial interferometers have the ability to constrain particle physics models at energy scales which remain inaccessible by any existing terrestrial accelerator.

Using the absence of a stochastic GWB signal in the data from the last three observing runs (O1-O3) of the LIGO/Virgo,  one can constrain $\Omega_{\rm gw}$ sourced by FOPT, and consequently test the underlying particle physics models. The O1-O3 data were analysed \cite{Romero:2021kby}  using either an approximated broken power-law (bpl) describing its main features in terms of the shape and the peak amplitude,  or a phenomenological model accounting for contributions from bubble collisions and sound waves. In the former approach, the stochastic  GWB spectrum is approximated by
\begin{equation}
    \Omega_{\rm bpl}(f)=\Omega_\star\left(\frac{f}{f_\star}\right)^{n_1}\left[1+\left(\frac{f}{f_\star}\right)^\Delta\right]^{(n_2-n_1)/\Delta}~,
\end{equation}
with $n_1=3$ from causality, $n_2$ equal to -4 or -1 for sound wave and for bubble collisions, respectively, and $\Delta$ equal to 2 or 4, for sound waves and approximating bubble collisions, respectively. One then runs a Bayesian search and model selection considering both a broken power law background and a CBC background approximated by a $f^{2/3}$ power law. In the second approach, it is adopted a phenomenological model with the dominant contribution coming either from bubble collisions (bc), following
\begin{equation}
    \Omega_{\rm bc}(f)h^2=1.67\times 10^{-5}\Delta\left(\frac{H_{\rm FOPT}}{\beta}\right)^2
    \left(\frac{\kappa_\phi \alpha}{1+\alpha}\right)^2 
    \left(\frac{100}{g_\star}\right)^{1/3} S_{\rm env}(f)~,
\end{equation}
or from sound waves (sw), following
\begin{eqnarray}
    \Omega_{\rm sw}(f)h^2=&&2.65\times 10^{-6}\left(\frac{H_{\rm FOPT}}{\beta}\right)\left(\frac{\kappa_{\rm sw} \alpha}{1+\alpha}\right)^2  \left(\frac{100}{g_\star}\right)^{1/3}\nonumber\\
    &&~~\times v_{\rm w}\left(\frac {f}{f_{\rm sw}}\right)^3
    \left(\frac{7}{4+3(f/f_{\rm sw})^2}\right)^{7/2}\Upsilon(\tau_{\rm sw})
\end{eqnarray}
where $h$ is the dimensionless Hubble parameter, $\Delta$ stands for the amplitude, $H_{\rm FOPT}$ is the Hubble parameter at $T_{\rm FOPT}$, $\beta$ is the inverse time duration of the transition, $\kappa_\phi$ denotes the fraction of vacuum energy converted into gradient energy of the scalar field, $\alpha$ is the energy released normalised by the radiation energy density, $g_\star$ is the number of relativistic degrees of freedom chosen to be 100, $S_{\rm env}$ is the spectral shape, $\kappa_{\rm sw}$
is the fraction of vacuum energy converted into the kinetic energy of the bulk flow, $v_{\rm w}$ is the bubble wall velocity, $f_{\rm sw}$ denotes the
present peak frequency, and $\Upsilon$
is a suppression factor due to the finite lifetime $\tau_{\rm SW}$.

The analysis following the generic power-law spectrum, led  to $95\%$ confidence level upper limits simultaneously on the GW energy density at 25 Hz from unresolved CBC and strong FOPT,  $\Omega_{\rm CBC}< 6.1\times 10^{-9}$ and  $\Omega_{\rm bpl}< 4.4\times 10^{-9}$, respectively \cite{Romero:2021kby} .
The analysis following the phenomenological model, constrained at 25 Hz the stochastic GWB to $\Omega_{\rm bc}<5.0\times 10^{-9}$ and $\Omega_{\rm sw}<5.8\times 10^{-9}$ at $95\%$ confidence level
due to bubble collisions and sound waves, respectively, 
\cite{Romero:2021kby} 

\subsection{Stochastic GWB: information on the early universe}

Several mechanisms in the early universe can create parity violation manifesting itself in the production of asymmetric amounts of right- and left- handed circularly polarised isotropic GWs. Since astrophysical sources of the GWB are unlikely to create circular polarisation, a detection of a circularly polarised signal may allow us to distinguish the cosmologically sourced component of the GWB from the astrophysically sourced one.  Analysing polarised GWB can then be used to constrain parity violating theories.

 Right- and left-handed correlators can then be written as
\begin{equation}
\begin{pmatrix}
    \langle h_{R}(f,\hat{\Omega})h_{R}^{*}(f',\hat{\Omega}') \rangle \\
    \langle h_{L}(f,\hat{\Omega})h_{L}^{*}(f',\hat{\Omega}') \rangle
    \end{pmatrix} \\
    = \frac{\delta(f-f')\delta^2(\hat{\Omega}-\hat{\Omega}')}{4\pi}
    \begin{pmatrix}
    I(f,\hat{\Omega}) + V(f,\hat{\Omega}) \\
    I(f,\hat{\Omega}) - V(f,\hat{\Omega})
    \end{pmatrix},
    \end{equation}
where $\langle\cdot\rangle$ represents the ensemble average and $I, V$ are the Stokes parameters, with $V$ characterising the asymmetry between right- and left-handed polarised waves, and $I (\geq |V|)$ the wave's total amplitude. Note that for $V=0$,  the above equation would just be the correlator for unpolarised isotropic GWB.

The cross-correlation estimator reads \cite{Romano:2016dpx}
\begin{eqnarray}
    \langle \hat{C}_{d_1 d_2} \rangle &=&\int_{-\infty}^{\infty}df\int_{-\infty}^{\infty}df'\delta_T(f-f')\langle s_{d_1}^{*}(f)s_{d_2}(f') \rangle \Tilde{Q}(f')\nonumber \\
    &=& \frac{3H_0^2 T}{10\pi^2}\int_0^{\infty}df\frac{\Omega'_{\rm gw}(f)\gamma_I^{d_1 d_2}(f)\Tilde{Q}(f)}{f^3},
    \label{eq:Y_estimator}
\end{eqnarray}
where
\begin{eqnarray}
    \Omega'_{\rm gw} &=& \Omega_{\rm gw}\bigg[1+\Pi(f)\frac{\gamma_V^{d_1 d_2}(f)}{\gamma_I^{d_1 d_2}(f)}\bigg],\nonumber \\
    \gamma_I^{d_1 d_2}(f) &=& \frac{5}{8\pi}\int d\hat{\Omega}(F_{d_1}^{+}F_{d_2}^{+*} + F_{d_1}^{\times}F_{d_2}^{\times*})e^{2\pi if\hat{\Omega}\cdot\Delta\Vec{x}}, \nonumber\\
    \gamma_V^{d_1 d_2}(f) &=& -\frac{5}{8\pi}\int d\hat{\Omega}(F_{d_1}^{+}F_{d_2}^{\times*} - F_{d_1}^{\times}F_{d_2}^{+*})e^{2\pi if\hat{\Omega}\cdot\Delta\Vec{x}},
    \label{eq:PVOmeg&ORF}
\end{eqnarray}
with $H_0$ the Hubble parameter, $T$ the observing time, $\delta_T(f) = \sin(\pi fT)/(\pi f)$, 
$\Tilde{s}_{d_1}(f)$ and $\Tilde{s}_{d_2}(f)$ the Fourier transforms of the strain time series of GW detectors $d_1$ and $d_2$, $\Tilde{Q}(f)$ the optimal filter taking into account detectors' strain power spectral densities, and $F_n^A = e_{ab}^A d_n^{ab}$ the contraction of the tensor modes of polarisation $A$ to the $n^{\rm th}$ detector's geometry. We denote by 
$\gamma_I^{d_1 d_2}$ 
the standard overlap reduction function of detectors $d_1, d_2$, and by 
$\gamma_V^{d_1 d_2}$ 
the overlap function associated with the parity violation term. The polarisation degree, $\Pi(f) = V(f)/I(f)$, can take any value between -1 (fully left polarisation) and 1 (fully right polarisation), with $\Pi = 0$ denoting an unpolarised isotropic GWB.

A strategy based on parameter estimation and subsequent fit of two GW models to O3 data, as well as simulated GW data (for A+ LIGO-Virgo detectors including KAGRA) was proposed in \cite{Martinovic:2021hzy}. The proposed method, using a hybrid frequentist-Bayesian approach, was employed in the search of
 a generic parity violation GW signal with a power-law behaviour, as well as a broken power-law spectral shape motivated by high energy physics.
The search has shown no evidence  for a parity violation signal in the most recent O3 data, while an upper limit of $\Omega (f_{\rm ref}=25 {\rm hz})=4.9\times 10^{-9}$
at $95\%$ confidence limit was imposed on the amplitude of a power-law GW model.
Simulating GW data for future detection prospects, and  considering a chiral turbulence source in the early universe it was demonstrated the importance of having a multi-detector network, while it was shown that even if we detect a turbulence signal, we may not be able to deduce its polarisation \cite{Martinovic:2021hzy}.

Subsequently, it was investigated \cite{Badger:2021enh} the ability of current or next generation detectors to detect parity violation sources from axion inflation focusing on the quadratic potential. 
It was shown \cite{Badger:2021enh} that using two 3g
detectors -- Einstein Telescope (ET) and Cosmic Explorer (CE) -- one may be  able to claim with at least
$95\%$ confidence a positive polarisation degree $\Pi$, while
a network of three 3g detectors (ET and 2 CEs) is needed
in order to make a confident claim about the detection
of a quadratic axion inflation signature.

\subsection{Stochastic GWB: information on dark matter (primordial black holes)}

Primordial black holes in the mass range $(30-100) M_\odot$ are interesting candidates for dark matter, as they occupy a narrow window between microlensing and cosmic microwave background constraints. 
By revisiting the primordial black holes merger rate to account for a more adequate description of black holes in their surrounding spacetime, it was shown that the merger rate constraints on the abundance of dark matter PBHs, imposed by LIGO, are evaded entirely \cite{Boehm:2020jwd,Boehm:2021kzq}.

Several studies in the literature examine whether some of the LIGO-Virgo detected CBC events are the outcome of PBH coalescences (see, for instance \cite{Sasaki:2016jop,Bird:2016dcv}). 
One of the mechanisms leading to PBH formation in the early universe is the large curvature fluctuations generated during an inflationary era. 
More precisely, PBHs result from the gravitational collapse of overdensed regions seeded by large curvature perturbations on small scales once the corresponding wavelength re-enters the horizon as the curvature perturbations exceed a critical value. Since the curvature perturbations couple to the tensor perturbations at second-order, such large curvature perturbations are accompanied by the scalar induced GWs in the radiation dominated era \cite{Matarrese:1997ay}.

The spectrum of the scalar induced GWB scales as the square of the curvature power spectrum ${\cal P}_\zeta (k)$, which at large scales is ${\cal O}(10^{-9})$ \cite{Planck:2018vyg}, set by cosmic microwave background data, and hence the scalar induced GWB remains unobservable by current or planned experiments. However, ${\cal P}_\zeta (k)$ may be considerable larger at small scales, reaching ${\cal P}_\zeta (k)\approx {\cal O}(0.01)$, and hence resulting to a GWB within reach of gravitational-wave observatories. 

A Bayesian search of scalar induces GWB in the data from Advanced LIGO and Virgo’s first,  second and third observing runs was performed in \cite{Romero-Rodriguez:2021aws}. Avoiding the choice of a particular inflationary model, the curvature power spectrum was parametrised by a well-localised peak depending on only three parameters:  its  position,  width  and  integrated  power, using a  log-normal  function \cite{Kohri:2018awv}
\begin{equation}
\mathcal{P}_\zeta(k) = \frac{A_\zeta}{\sqrt{2\pi} \Delta_\zeta} \exp\left[- \frac{\ln^2(k/k_*)}{2\Delta_\zeta^2} \right] \,,
\end{equation}
where $A_\zeta$ is the integrated power of the peak, $\Delta_\zeta$ determines its width and $k_*$ its position.

The scalar induced GW spectrum  is peaked around the same wavenumber as $\mathcal{P}_\zeta(k)$, corresponding to the frequency $f_*/{\rm Hz} = 1.6\times 10^{-15} k_*/{\rm Mpc}^{-1}$.
At $\Delta_\zeta\ll 1$, for which the amplitude of the induced GWs as well as the generated PBH abundance become independent of $\Delta_\zeta$, the
 peak amplitude of the GWB is $\Omega_{\rm gw} = \mathcal{O}(10^{-5}) A_\zeta^2$. Since the LIGO-Virgo detectors, are sensitive to frequencies around $f\gtrsim 10\,{\rm Hz}$,  they have the potential to probe peaks in $\mathcal{P}_\zeta(k)$, at scales larger than $10^{15}\,{\rm Mpc}^{-1}$, which re-enter the horizon at temperatures $\gtrsim 10^8$ Gev.

Performing a Bayesian analysis -- including also potential contributions from  CBC --  no evidence was found for a scalar induced GWB in the  LIGO-Virgo O1,  O2 and O3 data \cite{Romero-Rodriguez:2021aws}. However, depending on the width $\Delta_\zeta$ of the peak, $ 95\%$ confidence level upper limits were imposed on the integrated  power  of  the  peak $A_\zeta$,  as  a  function  of  its  position $k_\star$. 
The obtained constraints are stronger than the ones arising form big bang nucleosynthesis and cosmic microwave background observations in the range $\mathcal{O}(10^{15}) < k_*/{\rm Mpc}^{-1} < \mathcal{O}(10^{18})$, and reach $A_\zeta\simeq 0.02$ for a narrow peak at $k_*\simeq 10^{17}\,{\rm Mpc}^{-1}$.  It is worth noting the constraints from LIGO and Virgo, at their design sensitivity, and from the Einstein Telescope can compete with those related to the abundance of the formed primordial black holes  \cite{Romero-Rodriguez:2021aws}.

\subsection{Stochastic GWB: Tests of General Relativity}

Einstein's theory of General Relativity allows for the existence of only two gravitational-wave polarisation modes (the tensor plus and cross modes), while general metric theories of gravity may allow for up to four additional polarisations: two vector modes and two scalar modes (the breathing and longitudinal scalar modes). The detection of any of these additional polarisation modes would imply violation of General Relativity, while their non-detection may  experimentally constrain extended theories of gravity. However, the three- detector Advanced LIGO-Virgo network is generally unable to distinguish the polarisation of transient gravitational-wave signals, like those from binary black holes. The twin LIGO detectors (Livingston and Hanford) are nearly co-oriented, implying that Advanced LIGO is sensitive to only a single polarisation mode. Moreover, even if the twin LIGO detectors were more favorably-oriented, one needs generically a network of at least six detectors to uniquely determine the polarisation content of a gravitational-wave transient.

Observation of the stochastic GWB offers a way to directly measure generic gravitational-wave polarisations. 
It has been shown that a stochastic GWB dominated by alternative polarisation modes may be missed by current searches optimised only for tensor polarisations. A Bayesian method to detect a generically-polarised stochastic GWB 
 was proposed in \cite{Callister:2017ocg} and subsequently 
used in the LIGO/Virgo observing runs.
This (generic) method relies on the construction of two odds ratios: one to determine if a stochastic GWB has been detected, another one to quantify evidence for the presence of alternative polarisations in the background. Employing this method, it was shown \cite{Callister:2017ocg} that flat scalar-polarised backgrounds of amplitude $\Omega^{({\rm S})}(25 ~{\rm Hz})\approx 2\times 10^{-8}$ can be confidently identified as non-tensorial with Advanced LIGO.
Note that $\Omega_{\rm gw}$ does not necessarily have the interpretation of an energy density in modified theories of gravity; one may think of it as a measure of the strain power in each polarisation. 
It was also shown \cite{Callister:2017ocg} that after three years of observation at design sensitivity, Advanced LIGO will be able to limit the amplitudes of tensor, vector, and scalar polarisations to 
$\Omega^{{(\rm T})}_{\rm gw}(25 ~{\rm Hz})<1.6\times 10^{-10}$, 
$\Omega^{({\rm V})}_{\rm gw}(25 ~{\rm Hz})<2.0\times 10^{-10}$,
and
$\Omega^{({\rm S})}_{\rm gw}(25 ~{\rm Hz})<5.0\times 10^{-10}$, respectively, at $95\%$ credibility.
In the case that a stochastic GWB of mixed polarisation is detected, Advanced LIGO alone will not be able to  precisely determine the parameters of the tensor, vector, and/or scalar components simultaneously due to large degeneracies between modes.
It is worth noting that while addition of Advanced Virgo does not particularly increase our ability to detect or identify backgrounds of alternative polarisations,
it does significantly improve our parameter estimation on power-law backgrounds, hence breaking the degeneracies of the twin LIGO analysis.

So far no evidence for scalar or vector modes was found in the data from Advanced LIGO and Advanced Virgo. The non-detection of scalar and vector polarised stochastic GWB is consistent with predictions of Einstein's theory of General Relativity.
Using the log-uniform prior for $\Omega_{\rm ref}$ and a Gaussian prior on the spectral index for each polarisation, the upper limits
\begin{equation}
\Omega^{({\rm S})}(25 ~{\rm Hz})\leq 2.1\times 10^{-8}~,~
\Omega^{({\rm V})}(25 ~{\rm Hz})\leq 7.9\times 10^{-9}~,~
\Omega^{({\rm S})}(25 ~{\rm Hz})\leq 6.4\times 10^{-9}~,\nonumber
\end{equation}
at the $95\%$ credible level on $\Omega_{\rm ref}$  for scalar, vector, and tensor polarisations, respectively, were obtained \cite{KAGRA:2021kbb}.

\section{Gravitational-wave transients}

The LIGO/Virgo/KAGRA (LVK) Collaboration has published 
\cite{LIGOScientific:2018mvr,LIGOScientific:2020ibl,LIGOScientific:2021usb}
catalogues of Gravitational-Wave Transient Catalogues (GWTC) of compact binary coalescences  candidate events observed by Advanced LIGO and Advanced Virgo during the first, second, and third  observing runs.

The GW data analysis of the LVK Collaboration is first done in low-latency so that public alerts can be generated that subsequently trigger follow-up astronomical observations, and then in higher latency in order to create GW catalogues. To analyse GW data from the last (O3) observing run, five pipelines were used in real time: an unmodelled burst search, and four matched-filter  pipelines. The matched-filter method relies on having a model of the signal, as a function of the intrinsic (two individual component masses and two dimensionless spin vectors) and extrinsic (luminosity distance, right ascension  and declination, inclination between total angular momentum and line-of-sight, time of merger,  reference phase, and polarisation angle) physical parameters of the source. Using the search pipelines, a template bank of GW waveforms is created, covering the appropriate intrinsic parameter space; it is consequently used to filter against the data and produce signal-to-noise ratio time series.

\subsection{GW transients: tests of General Relativity}

Gravitational waves emitted by CBCs allow us to test GR in extreme environments of strong gravitational fields, large spacetime curvature, and velocities close to the speed of light.
Considering  the most significant signals, namely the ones with a false-alarm rate (FAR)
$<10^{-3} {\rm yr}^{-1}$, from the gravitational-wave transient catalogue GWTC02
\cite{LIGOScientific:2020ibl}, observational constraints of General Relativity (GR) were reported \cite{LIGOScientific:2020tif} by the LVK Collaboration, following a twofold approach. 
First, the consistency of GR predictions with the data was tested and then, by parametrising modifications to GW waveforms, the amount of allowed deviations from GR predictions was constrained.

To investigate whether the GR waveforms are successful in describing the data, one studies the residual strain, once the best-fit template for each event has been subtracted. Such an analysis has not found any statistically significant deviations between the observed residual power and the detector noise \cite{LIGOScientific:2020tif}.
Within GR the final state of the coalescence of two black holes will be a single perturbed Kerr black hole, gradually relaxing to its Kerr stationary state by emitting GWs which correspond to a specific set of characteristic quasinormal modes.
The explicit analysis of the GWTC02 events, has shown \cite{LIGOScientific:2020tif} that all GW transient signals were produced by merging Kerr black holes in agreement with Einstein’s theory of General Relativity.

In addition, one can use the GWTC02 events to constrain parameters of phenomenological modifications to General Relativity. In the context of GR, GWs far from their source propagate along null geodesics, with energy $E$ and momentum $p$ related by the dispersion relation
$E^2=p^2 c^2$, with $c$ denoting the speed of light. To investigate modifications to General Relativity, a phenomenological approach was considered introducing a modified dispersion relation \cite{LIGOScientific:2019fpa}
\begin{equation}
    E^2=p^2c^2+A_\alpha p^\alpha c^\alpha~, ~\mbox{with}~ \alpha=0, 0.5, 1.0, 1.5, 2.5, 3.0,3.5, 4.0~,
\end{equation}
with $A_\alpha$ and $\alpha$ two phenomenological parameters. Such an expression holds for a variety of proposed extensions to General Relativity. For instance, $\alpha=0,  A_\alpha>0$ represents massive gravity with graviton mass $m_{\rm g}=A_0^{1/2}c^{-2}$; note that General Relativity is recovered for $A_\alpha, \forall \alpha$.
All cases apart $\alpha=0$ correspond to a Lorentz-violating dispersion relation.

In frequency-domain, the strain in Fourier space, within General Relativity, is
\begin{equation}
    {\tilde h}(f)=A(f)e^{i\Phi(f)}~,
\end{equation}
A non-zero $A_\alpha$ will lead to a frequency-dependent dephasing of the GW signal, $\delta \Phi_\alpha(f)$, building up as the GW propagates towards Earth. For a given model (i.e., given the values of $A_\alpha, \alpha$) the dephasing $\delta \Phi_\alpha(f)$ depends on the binary’s luminosity distance,  the binary’s detector-frame  chirp mass, and  the effective wavelength parameter used in the sampling, defined in terms of binary’s redshift, and a distance parameter for a given cosmological model.

Using the gravitational-wave transient catalogue GWTC02
\cite{LIGOScientific:2020ibl}
$90\%$ credible upper bounds on the absolute value of the modified dispersion relation parameter $A_\alpha$ as a function of $\alpha$ where obtained in \cite{LIGOScientific:2020tif}. The analysis has shown no evidence for GW dispersion, constraining the Lorentz-violating dispersion parameters. The obtained constraint on the graviton mass is \cite{LIGOScientific:2020tif}
\begin{equation}
    m_{\rm g}\leq 1.76\times 10^{-23} {\rm eV}/{\rm c}^2~,
\end{equation}
with $90\%$ credibility; it is an improvement of 1.8 over Solar System bounds.

Since we have only 3 active detectors, we cannot simultaneously constrain all six possible GW polarisations allowed in generic metric theories of gravity.
For GW transients, one can then compare the likelihood of having purely scalar or purely vector polarisations against the pure tensor case, predicted by General Relativity.
Such an approach, using multiple binary black hole signals reported in the gravitational-wave transient catalogue GWTC02
\cite{LIGOScientific:2020ibl}, has shown \cite{LIGOScientific:2020tif} no evidence in favor of non-tensor GWs.

\subsection{GW transients: tests of modified gravity models}

Several modified gravity phenomenological models have been proposed in the literature, in an attempt to find an explanation for the dark energy. Gravitational waves offer the means to test/constrain such modified gravity models by studying the propagation of GWs across cosmological distances \cite{Lambiase:2020vul,LISACosmologyWorkingGroup:2019mwx}.

In the context of General Relativity, gravitational waves travelling on a four-dimensional Friedmann-Lema\^{i}tre-Roberson-Walker background, obey the linearised evolution equation
\begin{equation}
    h_A^{''}+2{\cal H}h_A^{'}+k^2h_A=\Pi_A~,
\end{equation}
where $A=+,\times$ stands for the two polarisation plus and cross modes, primes denote derivatives with respect to conformal time $\eta$, related to the cosmological time through ${\rm d}\eta={\rm d}t/a(t)$ with $a(t)$ the scale factor, ${\cal H}$ is the Hubble parameter in conformal time $\eta$, and $\Pi_A$ denotes the source term related to the anisotropic stress tensor.
The GW propagation equation above, gets modified in a generic modified gravity model into
\begin{equation}
 h_A^{''}+2[1-\delta(\eta)]{\cal H}h_A^{'}+[c_{\rm T}^2(\eta)k^2+m_{\rm T}^2(\eta)]k^2h_A=\Pi_A~,   
\end{equation}
where three new quantities have been introduced.
The function $\delta(\eta)$ modifies the friction term and hence affects the amplitude of a GW propagating across cosmological distances.
The tensor velocity $c_{\rm T}$ can be in general time and scale dependent; in General Relativity it is equal to the speed of light $c$.
The mass of the tensor mode $m_{\rm T}$, can be non-zero in the context of a modified gravity theory. These three quantities are in principle testable with GW data.

The modification in the tensor sector leads to the gravitational-wave  luminosity distance $d_{\rm L}^{({\rm gw})}(z)$, which is different from the standard electromagnetic luminosity distance
\begin{equation}
    d_{\rm L}^{(\rm em)}(z)=(1+z)\int_0^z\frac{{\rm d}{\tilde z}}{H({\tilde z})}~,
\end{equation}
where 
\begin{equation}
    H(z)=H_0\sqrt{\Omega_{\rm m}(1+z)^3+\Omega_{\rm r}(1+z)^4+\Omega_{\rm DE}(z)}
    ~,\nonumber
\end{equation}
with $\Omega_{\rm m}, \Omega_{\rm r}$ and $\Omega_{\rm DE}$ the present matter density, radiation density and DE fractions, respectively. A simple 
phenomenological parametrisation 
\begin{equation}
\Xi(z)\equiv\frac{d_{\rm L}^{(\rm gw)}(z)}{d_{\rm L}^{(\rm em)}(z)}=\Xi_0+\frac{1-\Xi_0}{(1+z)^n}~,
\end{equation}
which depends on the (positive) parameters $\Xi_0$ and $n$ (with  $\Xi_0=1$ in General Relativity) has been proposed in \cite{Belgacem:2018lbp}.
This approach has been employed in \cite{LISACosmologyWorkingGroup:2019mwx} to test several modified gravity models -- Horndeski or the more general degenerate higher order scalar-tensor theories, non-local infrared modifications of gravity, bi-gravity theories, as well as theories with extra or varying dimensions -- using standard sirens at GW detectors such as LISA or third-generation ground based experiments.
More precisely, constructing mock source catalogues and performing Markov Chain Monte Carlo studies of the likelihood obtained from LISA standard sirens combined  with other cosmological datasets, it was found \cite{LISACosmologyWorkingGroup:2019mwx} that the parameter $\Xi_0$ can be measured to the percent level accuracy. 

In conclusion, it has been shown that the modified GW propagation is an important observable for advanced GW detectors; supermassibe black hole mergers binaries detectable with LISA can provide a powerful probe of modified gravity and dark energy.


\subsection{GW transients: information on dark matter (axion-like particles)}

Pseudo-scalar (axion-like) particles are hypothetical scalar particles that appear in many fundamental theories.
String theory compactifications generally predict a number of light axions; they represent  one of the most compelling candidates for dark matter.

Consider axions that couple to nuclear matter in the same way as the QCD axion, but with masses that are relatively light. 
Such axions can be sourced by compact objects with a high nuclear density,  for instance neutron stars. 
If the axion decay constant is $f_{\rm a}\lesssim 10^{18}$, neutron stars with radius of the order of 10 km will develop an axion profile, interpolating from $\pm \pi f_{\rm a}$ near the neutron star surface to 0 at spatial infinity. For two neutron stars, with each one  within the axion's Compton wavelength, $\lambda_{\rm a}\equiv 1/m_{\rm a}$,  the axion field mediates an additional force, which at leading order is
\begin{equation}
{\bf F}_{\rm a}=-\frac{Q_1Q_2}{4\pi r^2}(1+m_{\rm a}r)\exp[-m_{\rm a} r]{\hat {\bf r}}~;
\end{equation}
$Q_{1,2}$ denote the axion charge carried by each neutron star, related to its radius $R_{1,2}$ through
\begin{equation}
Q_{1,2}=\pm 4\pi^2f_{\rm a}R_{1,2}~.
\end{equation}
The axion force ${\bf F}_{\rm a}$ is attractive or repulsive, depending on whether the axion field values have the same or opposite sign, respectively, on the surfaces of the two neutron stars.
Compact objects with a scalar charge may have important implications for binary neutron stars or neutron star-black hole mergers, leading to the emission of axion radiation and an axion-mediated fifth force, which effects the orbital dynamics leaving potentially detectable fingerprints in the inspiral GW waveform
 \cite{Sagunski:2017nzb}.
 
More precisely, if such neuron stars form a binary system, the axion field might also radiate axion waves during the compact objects coalescence. Considering circular orbits, the radiation power at leading order reads
\begin{equation}
P_{\rm a}=\frac{(Q_1M_2-Q_2M_1)^2}{12\pi(M_1+M_2)^2} r^2\omega^4\left(1-\frac{m_{\rm a}^2}{\omega^2}\right)^{3/2}~,
\end{equation}
where $\omega$ denotes the orbital frequency and $r$ the separation between the two neutron stars of masses $M_1, M_2$. The axion radiation turns on once the orbital frequency exceeds the axion mass.

Using an effective field theory approach, the first post-Newtonian corrections to the orbital dynamics, radiated power, and gravitational waveform for binary neutron star mergers in the presence of an axion have been calculated in \cite{Huang:2018pbu}. It was shown that Advanced LIGO at design sensitivity, can potentially exclude axions wuth mass $m_{\rm a}\lesssim 10^{-11} eV$  and decay constant
$f_{\rm a}\sim (10^{14}-10^{17})$ GeV.

Subsequently, a search for such axions has been conducted \cite{Zhang:2021mks} by analysing the gravitational waves from the binary neutron star inspiral GW170817.
Let us write the frequency domain waveform as \cite{Zhang:2021mks} 
\begin{equation}
    h(f)\simeq H(f)\exp[i\Psi(f)]~,
\end{equation}
with
\begin{equation}
    \Psi=\Psi_{\rm GR}+\Psi_{\rm a}+{\cal O}(Q_{1,2}^4)+{\cal O}(Q_{1,2}^2v^2)~,
\end{equation}
where $\Psi_{\rm GR}$ is the phase of the  waveform within General Relativity, $\Psi_{\rm a}$ denotes the leading order phase correction due to the axion field, and $v^2$ counts the post-Newtonian order.

The effects of the axion field in the waveform template can be parametrised by the axion mass $m_{\rm a}$ and a  dimensionless parameter $\gamma_{\rm a}\equiv Q_1Q_2/(4\pi G M_1M_2)$ that characterises the relative strength of the axion and gravitational forces between the two neutron stars. Performing a Bayesian analysis of GW17081, the posterior distribution over the waveform parameters suggests no significant evidence for such axion fields \cite{Zhang:2021mks}. The null result constrainted  axions with masses below $10^{-11}$ eV by excluding the ones with decay constants ranging from $1.6\times 10^{16}$ GeV to $ 10^{18}$ GeV at a $3 \sigma$ confidence level. This is the first constraint on axions imposed from neutron star inspirals \cite{Zhang:2021mks}.

It is worth noting that such constraint may be improved if the SNR of the merger event is enhanced; the constraint on $f_{\rm a}$ will improve by $\sqrt N$ if the SNR is improved by $N$. The constraint could also be improved by roughly 2 orders of magnitude by considering a neutron star-black hole merger.

\subsection{GW transients: information on quantum gravity theories}

Gravitational waves can also provide means to test quantum gravity (QG) theories, challenging one's naive claim that any late-time QG imprint will be Planck-suppressed and hence undetectable.
The reason being the existence of a long-range non-perturbative mechanism, called dimensional flow (scale-dependence of the dimensionality of spacetime), shared by most QG candidate theories.  

Let us consider propagation of gravitational waves in the context of QG theories. Quantisation of spacetime geometry introduces two types of change relevant for the propagation of GWs: an anomalous spacetime measure  and a kinetic operator. The former captures the way that volumes scale and the latter is related to the modified dispersion relations. 
Given a spacetime metric, a kinetic operator and a compact source, the GW amplitude, in radial coordiantes and in the local wave zone, reads \cite{Calcagni:2019kzo,Calcagni:2019ngc} 
\begin{equation}
    h(t,r)\sim f_h(t,r)\left(\frac{\ell_\star}{r}\right)^\Gamma~, ~ [f_h]=0~,
\end{equation}
where $\Gamma\simeq d_{\rm H}/2-d_{\rm H}^k/d_{\rm s}$, with
$d_{\rm H}, d_{\rm H}^k$ and $d_{\rm s}$ the Hausdorff dimension, the Hausdorff dimension in momentum space, and the spectral dimension, respectively. 
Clearly, $h$ is the product of a dimensionless function $f_h$, depending on the source  and  the type of correlation function, and a power-law distance behaviour.
This is a rather general feature in quantum gravity theories.

The amplitude of GWs propagating in such a background is \cite{Calcagni:2019kzo,Calcagni:2019ngc} 
\begin{equation}
    h\propto \frac{1}{d_{\rm L}^{({\rm gw})}}~~,~~ \frac{d_{\rm L}^{(\rm gw)}}{d_{\rm L}^{(\rm em)}}=1+\epsilon\left(\frac{d_{\rm L}^{({\rm em})}}{\ell_\star}\right)^{\gamma-1}~,
    \label{dim-flow}
\end{equation}
with $\epsilon={\cal O}(1)$, $\gamma\neq 1$, and $\ell_\star$ a characteristic scale of
the geometry. If there is only one fundamental length scale $\ell_\star={\cal O}(\ell_{\rm Pl})$, the above equation relating the GW luminosity distance to the electromagnetic one, is exact and $\gamma =\Gamma_{\rm UV}$, namely scales at which QG corrections are important. If however $\ell_\star$ is a mesoscopic scale $\Gamma_{\rm meso}$, denoting intermediate scales where the corrections are small but non-negligible, then the above equation is valid only near the infrared regime and $\gamma=\Gamma_{\rm meso}\approx 1$. The values of $\Gamma_{\rm UV}$ and $\Gamma_{\rm meso}$ are known for large variety of QG candidate theories (including group field theory, spin foams, loop quantum gravity, causal dynamical triangulations, $\kappa$-Minkowski, low-energy limit of string theory, asymptotic safety).

The left-hand side of Eq.~(\ref{dim-flow}) is the strain measured in a GW
interferometer, while the right-hand side gives the luminosity distance measured for the optical counterpart of the standard siren.
GW observations can thus be used to constraint $\ell_\star$ and $\gamma$, by constraining the
ratio ${d_{\rm L}^{(\rm gw)}(z)/d_{\rm L}^{(\rm em)}}(z)$ as a function of the redshift $z$ of the source. 

Considering  two examples of multimessenger standard sirens, namely the binary neutron-star merger GW170817 and a simulated supermassive black hole merger event detectable with LISA, it was found \cite{Calcagni:2019kzo} that only group field theory, spin foams or loop quantum gravity could generate a signal detectable with standard sirens.

\section{Conclusions}

Gravitational waves give us the means to unravel mysteries of our universe, allowing us to test theories describing its early stages, when it was opaque to electromagnetic radiation. Using direct observations  of gravitational waves we can now test particle physics models at energy scales which are not accessible to any foreseeable particle accelerator, test Einstein's theory of General Relativity in the strong gravity regime, and also constrain modified/extended gravity models.  Gravitational waves offer us an independent way to study the large scale structure of our universe, and to obtain a useful insight on astrophysical properties of compact objects. Finally,  they allow us to constrain some of the most promising dark matter candidates, while  they open the way to test quantum gravity  theories. 
 

\end{document}